\documentclass[aps,prb,twocolumn,showpacs,superscriptaddress]{revtex4-1}  % for review and submission
\usepackage{graphicx}  % needed for figures
\usepackage{dcolumn}   % needed for some tables
\usepackage{bm}        % for math
\usepackage{amssymb}   % for math
\usepackage{color}
\usepackage{physics}
\usepackage[colorlinks = true,
            linkcolor = blue,
            urlcolor  = blue,
            citecolor = blue,
            anchorcolor = blue]{hyperref}
\usepackage[all]{hypcap}

\usepackage[applemac]{inputenc}
\usepackage{amsmath}
\usepackage{mathptmx}
\newcommand{\refsub}[2][{}]{\hyperref[#2]{\ref{#2}(#1)}} 

\newcommand{\cpi}{CePt$_2$In$_7$}

\begin{document}

\title[Investigation of the commensurate magnetic structure in heavy fermion \cpi]{Investigation of the commensurate magnetic structure in heavy fermion \cpi\ using magnetic resonant X-ray~diffraction}

\author{Nicolas Gauthier}  
\email{nicolas.gauthier4@gmail.com}
\affiliation{Laboratory for Scientific Developments and Novel Materials, Paul Scherrer Institut, 5232 Villigen, Switzerland}
\author{Didier Wermeille}  
\affiliation{XMaS, UK-CRG, European Synchrotron Radiation Facility, BP220, F-38043 Grenoble Cedex, France}
\affiliation{Oliver Lodge Laboratory, Department of Physics, University of Liverpool, Oxford Street, Liverpool L69 7ZE, United Kingdom}
\author{Nicola Casati}  
\affiliation{Laboratory for Synchrotron Radiation, Paul Scherrer Institut, 5232 Villigen, Switzerland}
\author{Hironori Sakai}
\affiliation{Advanced Science Research Center, Japan Atomic Energy Agency, Tokai, Ibaraki, 319-1195, Japan}
\author{Ryan E. Baumbach}  
\altaffiliation[Present address: ]{Condensed Matter Group, National High Magnetic Field Laboratory, Florida State University, Tallahassee, Florida 32310, USA}
\affiliation{Condensed Matter and Magnet Science, Los Alamos National Laboratory, Los Alamos, New Mexico 87545, USA}
\author{Eric D. Bauer}  
\affiliation{Condensed Matter and Magnet Science, Los Alamos National Laboratory, Los Alamos, New Mexico 87545, USA}
\author{Jonathan S. White}  
\email{jonathan.white@psi.ch}
\affiliation{Laboratory for Neutron Scattering and Imaging, Paul Scherrer Institut, 5232 Villigen, Switzerland}

\begin{abstract}
We investigated the magnetic structure of the heavy fermion compound \cpi\ below $T_N~=~5.34(2)$~K using magnetic resonant X-ray diffraction at ambient pressure. The magnetic order is characterized by a commensurate propagation vector $\bm{k}_{1/2}~=~\left( \frac{1}{2} , \frac{1}{2}, \frac{1}{2}\right)$ with spins lying in the basal plane. {Our measurements did not reveal the presence of an incommensurate order propagating along the high symmetry directions in reciprocal space but cannot exclude other incommensurate modulations or weak scattering intensities.} The observed commensurate order can be described equivalently by either a \mbox{single-$\bm{k}$} structure or by a multi-$\bm{k}$ structure. Furthermore we explain how a commensurate-only ordering may explain the broad distribution of internal fields observed in nuclear quadrupolar resonance experiments (Sakai \textit{et al.} 2011, \textit{Phys. Rev. B} \textbf{83} 140408) that was previously attributed to an incommensurate order. We also report powder X-ray diffraction showing that the crystallographic structure of \cpi\ changes monotonically with pressure up to $P~=~7.3$~GPa at room temperature. The determined bulk modulus $B_0~=~81.1(3)$~GPa is similar to the ones of the Ce-115 family. Broad diffraction peaks confirm the presence of pronounced strain in polycrystalline samples of \cpi. We discuss how strain effects can lead to different electronic and magnetic properties between polycrystalline and single crystal samples.
\end{abstract}

\pacs{75.25.-j, 75.30.Mb, 61.05.cp}

%\noindent{\it Keywords\/}: Heavy fermions, resonant magnetic X-ray diffraction, magnetic structure,  superconductivity

\maketitle

%%%%%%%%%%%%%%%%%%%%%%%%%%%%%%%%%%%%%%%%%%%
%%%%%%%%%%%%%%%%%%%%%%%%%%%%%%%%%%%%%%%%%%%
%%%%%                INTRODUCTION
%%%%%%%%%%%%%%%%%%%%%%%%%%%%%%%%%%%%%%%%%%%
%%%%%%%%%%%%%%%%%%%%%%%%%%%%%%%%%%%%%%%%%%%
\section{Introduction}
Electrons can gain a large effective mass due to strong electronic correlations in crystals. Such materials are referred to as heavy fermion compounds and often have complex phase diagram due to the interplay of spin and electronic degrees of freedom. Of particular interest are the Ce-115 compounds Ce\textit{M}In$_5$ (\textit{M}~=~Co, Rh, Ir) that have been investigated for more than 15 years and yet their properties are still not completely understood.\cite{Thompson2012} These materials, which offer a unique playground to study quantum criticality,\cite{Gegenwart2008} are part of the larger family Ce$_n$\textit{M}$_m$In$_{3n+2m}$ (\textit{M}~=~Co, Rh, Ir, Pd, Pt) derived from the simple cubic CeIn$_3$: they are formed from CeIn$_3$ layers separated by \textit{M}In$_2$ layers. This separation of the Ce planes makes them generally more two-dimensional (2D) relative to the three-dimensional (3D) cubic CeIn$_3$. 
Furthermore, the hybridization of the Ce 4$f$-electrons with the conduction electron bands is controlled by the local environment of the In and \textit{M} atoms.\cite{Willers2015,Haule2010}
It is therefore possible to investigate the effects of the dimensionality and the hybridization strength on the interplay between magnetism and superconductivity in these compounds. 

\cpi\ is a member of this family with $n~=~1$ and $m~=~2$. It is closely related to the Ce-115s and is obtained by adding a second \textit{M}In$_2$ plane in between the CeIn$_3$ planes. This larger separation of the planes containing Ce suggest that this system is more 2D than the Ce-115s. \cpi\ crystallizes in a body-centered tetragonal structure with space group \textit{I}4/\textit{mmm} and the magnetic Ce ion sits at the  Wyckoff \textit{2b} positions.\cite{Klimczuk2014,Sakai2014} It has an antiferromagnetic (AFM) order with $T_N\approx 5.5$~K at ambient pressure. This order is suppressed with pressure and a superconductivity dome emerges around the AFM quantum critical point (QCP), with a maximum $T_c~=~2.1$~K near the critical pressure $P_c\approx3.4$~GPa, and which is also where an effective mass enhancement is observed.\cite{Bauer2010a} This phase diagram is very similar to the analogous compound CeRhIn$_5$,\cite{Park2006} which is often described as a two-dimensional analogue of CeIn$_3$. Quantum oscillations reveal that the microscopic electronic properties of \cpi\ are more closely related to CeIn$_3$ than CeRhIn$_5$, indicating that \cpi\ is a better 2D analog of CeIn$_3$.\cite{Altarawneh2011} The 2D nature of the electronic properties is also suggested by specific heat measurements.\cite{Krupko2016} Optical measurements indicate a hybridization strength in \cpi\ similar to the one in CeIn$_3$ and CeRhIn$_5$.\cite{Chen2016a}

Nuclear quadrupolar resonance (NQR) measurements revealed the presence of two characteristic pressures in \cpi.\cite{Sakai2014} The first one at $P^*~=~2.4$~GPa corresponds to a transition from localized to itinerant Ce 4$f$-electrons. The second one at $P_c\approx3.4$~GPa corresponds to the AFM QCP. In CeRhIn$_5$, these characteristic pressures are very close to each other and it was suggested that the superconductivity emerges from the Kondo breakdown QCP.\cite{Park2011}
Indeed, recent theoretical work proposes an enhancement of singlet superconductivity near a Kondo breakdown QCP,\cite{Pixley2015} which may explain the behaviour of \cpi\ and CeRhIn$_5$.\cite{Sakai2014,Park2011} The detailed understanding of \cpi\ also requires an accurate description of its magnetic order at ambient pressure and its evolution (or stability) under pressure. However, up until now only limited details of the nature of the magnetic order have been reported. NQR measurements on polycrystalline samples indicate a commensurate order and suggest a propagation vector $(\frac{1}{2},\frac{1}{2})$ in the basal plane.\cite{Aproberts-Warren2010} On the other hand, the results obtained using the same technique applied to single crystals were interpreted in terms of a coexistence of commensurate and incommensurate orders.\cite{Sakai2011} From muon spin rotation measurements, a commensurate order was proposed for polycrystalline samples.\cite{Mansson2014} A possible reason for these discrepancies is that the inherently larger surface strain of grains in polycrystalline samples provides a means to enhance the stability of the commensurate order.\cite{Sakai2011} It was also observed that the superconducting dome is broader for powders than for single crystals, suggesting a commensurate order to be more favourable for superconductivity.\cite{Sidorov2013} However, both direct measurements of the magnetic order and its propagation, and evidence for the proposed crystallographic strain in powder samples are yet to be reported.

Neutron scattering could clarify the bulk magnetic structure but it is challenging for \cpi\ because of the generally small size of single crystals, the large neutron absorption cross-section by In, and the small expected moment size. These limitations can be overcome by using magnetic resonant X-ray diffraction (MRXD) as an alternative scattering technique for determining the magnetic structure. We performed MRXD measurements on \cpi\ and we report here a model for the magnetic order at $T~=~1.8$~K and ambient pressure. We also report the pressure dependence of its crystallographic structure at room temperature up to $P~=~7.3$~GPa, which changes monotonically in the range of applied pressure.

%%%%%%%%%%%%%%%%%%%%%%%%%%%%%%%%%%%%%%%%%%%
%%%%%%%%%%%%%%%%%%%%%%%%%%%%%%%%%%%%%%%%%%%
%%%%%                Experimental details
%%%%%%%%%%%%%%%%%%%%%%%%%%%%%%%%%%%%%%%%%%%
%%%%%%%%%%%%%%%%%%%%%%%%%%%%%%%%%%%%%%%%%%%
\section{Experimental details}
High purity single crystals of \cpi\ were synthesized as described previously.\cite{Tobash2012} The 0.38~mg sample used for the MRXD experiment was characterized by specific heat and magnetic susceptibility using a Quantum Design PPMS and MPMS, respectively. The results are in good agreement with the previously reported measurements.\cite{Tobash2012} The long range magnetic order is observed from a sharp peak in the specific heat at $T_N~=~5.36(2)$~K [Fig.~\refsub[c]{fig2}] and the high purity of the sample is indicated by the absence of other peaks, compared to previous reports.\cite{Tobash2012,Bauer2010} For the MRXD experiment, the plate-like sample with the $c$-axis perpendicular to the plate was fixed on a copper holder with silver Electrodag 1415 and mounted in a Joule Thomson cryostat on the bending magnet XMaS beamline, at the ESRF. The measurements were carried out using a Vortex Si Drift Diode detector. The (220) reflection of a LiF analyser crystal was used for the polarization analysis measurements. Except for photon energy dependent scans, all the measurements were carried out at $E~=~6.166$~keV, the Ce-$L_{II}$ absorption edge. 
The azimuthal scans presented in Fig.~\ref{fig3} were corrected for X-ray absorption. The absorption correction was calculated by a finite element analysis assuming an absorption coefficient $\mu~=~436.425$~mm$^{-1}$ for \cpi, a beam size of $0.7\times0.3$~mm$^2$ and a sample size of $0.79\times0.62\times0.02$~mm$^3$. The accuracy of this correction for the magnetic peaks was verified by comparison with azimuthal scans measured on structural peaks. 

Powder X-ray diffraction measurements under hydrostatic pressure were performed at the MS-X04SA beamline, Swiss Light Source at the Paul Scherrer Institut.\cite{Willmott2013} A 2D Pilatus 6M detector was used. LaB$_6$ was used as a standard for calibration of the detector position as well as the instrumental parameters. Single crystals of \cpi\ were finely ground, mixed with quartz powder and loaded in a diamond anvil pressure cell using methanol:ethanol 4:1 as a pressure medium. Quartz was used as an in-situ pressure calibrant.\cite{Angel1997} Measurements were performed with a photon wavelength $\lambda~=~0.56491~\mathrm{\AA}$ in the angular range $1^\circ<2\theta<35^\circ$ at room temperature ($T~=~293$~K) up to a maximal pressure $P~=~7.3$~GPa. The data reduction was performed with the Dioptas software\cite{Prescher2015} and \textsc{FullProf} was used for Rietveld refinement of the one-dimensional diffraction patterns.\cite{Fullprof}

%%%%%%%%%%%%%%%%%%%%%%%%%%%%%%%%%%%%%%%%%%%%
%%%%%%%%%%%%%%%%%%%%%%%%%%%%%%%%%%%%%%%%%%%%
%%%%%%                Experimentals results 
%%%%%%%%%%%%%%%%%%%%%%%%%%%%%%%%%%%%%%%%%%%%
%%%%%%%%%%%%%%%%%%%%%%%%%%%%%%%%%%%%%%%%%%%%
\section{Experimental results}

\subsection{Magnetic Resonant X-ray Diffraction}
\label{secMRXD}

The magnetic order of \cpi\ was successfully observed using MRXD, revealing unambiguously its commensurate propagation vector. Bragg peaks consistent with a propagation vector $\bm{k}_{1/2}~=~\left( \frac{1}{2} , \frac{1}{2}, \frac{1}{2}\right)$ were observed at $T~=~1.8$~K. The magnetic origin of these Bragg peaks was verified by the resonance at the Ce-$L_{II}$ absorption edge as well as polarization analysis. $\bm{Q}$-scans around the magnetic Bragg peak $\bm{Q}~=~(0.5, -0.5, 6.5)$ are presented in Figs.~\refsub[a]{fig1}-\refsub[c]{fig1} at $T~=~1.8$~K and can be compared with background scans done at 10~K. This magnetic Bragg peak has the same widths and shapes along $H$, $K$ and $L$ as the structural Bragg peak $\bm{Q}~=~(1,-1,6)$. This indicates that the magnetic peak widths are limited by the crystal mosaicity and that a 3D long range magnetic order is achieved. Several other peaks consistent with $\bm{k}_{1/2}$ were measured. It was observed that all experimentally accessible magnetic Bragg peaks have non-zero intensity, indicating the absence of any selection rules of the magnetic structure. 

\begin{figure}[!htb]
\centering
\includegraphics[scale=1.05]{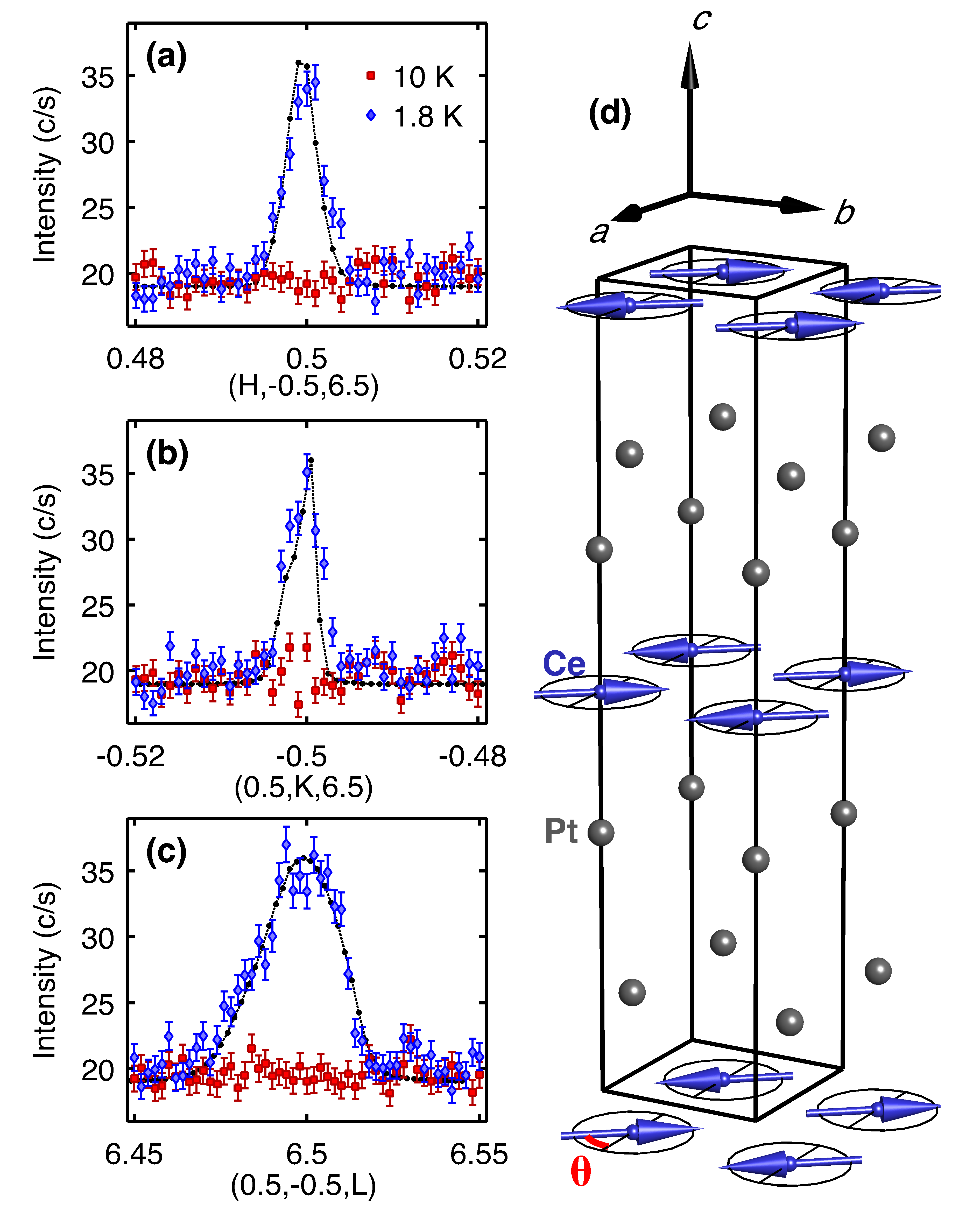}
\caption{(a) $H$-scans, (b) $K$-scans and (c) $L$-scans around the magnetic Bragg peak $\bm{Q}~=~(0.5, -0.5, 6.5)$ in the ordered state ($T~=~1.8$~K) and in the paramagnetic state ($T~=~10$~K). The connected black dots are corresponding scans around the structural Bragg peak $\bm{Q}~=~(1,-1,6)$, and scaled to provide a comparison between the peak widths and shapes. (d) Single-$\bm{k}$ magnetic structure of \cpi\ with $\bm{k}_{1/2}~=~\left( \frac{1}{2} , \frac{1}{2}, \frac{1}{2}\right)$ and moments aligned in the basal plane at an angle $\theta$ from the $a$-axis. Indium atoms have been omitted for clarity.}
\label{fig1}
\end{figure}

The fluorescence intensity of the sample was measured as function of the incident photon energy. It shows a maximum around $E_i~=~6.167$~keV corresponding to the Ce-$L_{II}$ absorption edge [Fig.~\refsub[a]{fig2}]. The intensity of the magnetic Bragg peak $\bm{Q}~=~(0.5, -0.5, 6.5)$ is strongly enhanced around this edge, indicating a resonant magnetic effect.\cite{Hill1996} In contrast, the intensity of the structural Bragg peak $\bm{Q}~=~(1,-1,6)$ shows a dip near this edge due to a larger absorption cross-section. The magnetic nature of the Bragg peak $\bm{Q}~=~(0.5, -0.5, 6.5)$ is further confirmed by the polarization analysis. The polarization $\sigma$ is defined to be perpendicular to the scattering plane and the polarization $\pi$ is parallel to it.\cite{Hill1996} In the electric dipole approximation of MRXD, charge scattering, related to the crystallographic structure, is allowed in the $\sigma-\sigma'$ channel and is forbidden in the $\sigma-\pi'$ channel. Magnetic scattering has the opposite behaviour and appears in the $\sigma-\pi'$ channel and not in the $\sigma-\sigma'$ one.\cite{Hill1996} The Bragg peak $\bm{Q}~=~(0.5, -0.5, 6.5)$ is present in the $\sigma-\pi'$ channel and absent in the $\sigma-\sigma'$ channel, clearly showing its magnetic nature [Fig.~\refsub[b]{fig2}]. This observation combined with the peak resonance at the Ce-$L_{II}$ edge establish unambiguously the magnetic origin of the Bragg peaks with the propagation vector $\bm{k}_{1/2}~=~\left( \frac{1}{2} , \frac{1}{2}, \frac{1}{2}\right)$. 

\begin{figure}[!htb]
\centering
\includegraphics[scale=1.05]{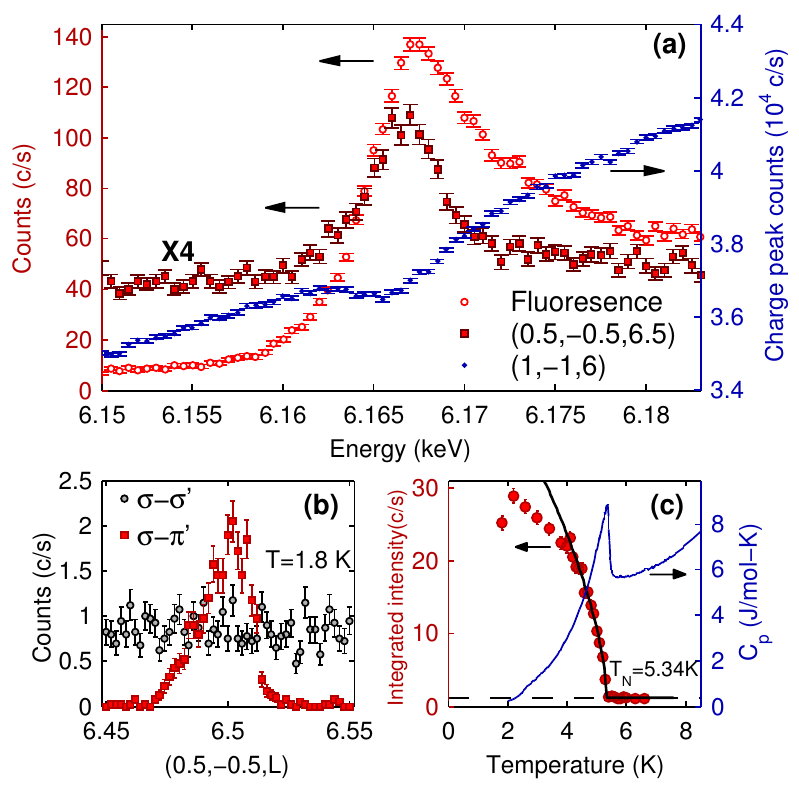}
\caption{(a) Photon energy scans near the $L_{II}$ absorption edge of Ce: fluorescence of the sample, resonance of the magnetic Bragg peak $\bm{Q}~=~(0.5, -0.5, 6.5)$ at the absorption edge and dip of the structural Bragg peak $\bm{Q}~=~(1,-1,6)$. (b) $L$-scan around the Bragg peak $\bm{Q}~=~(0.5, -0.5, 6.5)$ with polarization analysis, showing that all the signal is in the $\sigma-\pi'$ channel and therefore magnetic. The finite signal in the $\sigma-\sigma'$ channel is due to nonmagnetic background. (c) Temperature dependence of the integrated intensity of the magnetic Bragg peak $\bm{Q}~=~(0.5, -0.5, 6.5)$, showing the transition at $T_N~=~5.34(2)$~K in agreement with the sharp peak in specific heat. The black line is a power-law fit to extract $T_N$. A beam injection occurred at $T~=~4.1$~K and the intensity below and above this value can not be compared directly.}
\label{fig2}
\end{figure}

The temperature dependence of the magnetic Bragg peak $\bm{Q}~=~(0.5, -0.5, 6.5)$ has been measured from $T~=~1.8$~K up to 7~K in the $\sigma-\pi'$ channel. The width and position of this peak are temperature independent from $T~=~1.8$~K to $T_N$. The integrated intensity indicates a N\'eel temperature of $T_N~=~5.34(2)$~K, as determined by a power law fit above 4.4~K [Fig.~\refsub[c]{fig2}]. This transition temperature is in good agreement with the sharp peak observed in specific heat. The obtained critical exponent $\beta~=~0.31(4)$ corresponds to a 3D Ising model with $\beta~=~0.326$ or a 3D XY model with $\beta~=~0.345$.\cite{Collins1989} Note that a beam injection occurred during the measurements at $T~=~4.1$~K and that the intensity above and below this temperature can not be compared accurately. However, a previous temperature dependence of the Bragg peak $\bm{Q}~=~(0.5, -0.5, 6.5)$ without the polarization analysis (not shown) does not have any feature at $T\approx4$~K.

The magnetic structure of the propagation vector $\bm{k}_{1/2}~=~\left( \frac{1}{2} , \frac{1}{2}, \frac{1}{2}\right)$ was determined with the help of representation analysis performed with \textsc{BasIreps}.\cite{Fullprof} Only two irreducible representations with non-zero basis functions are possible at the Ce position $(0,0,0.5)$ in the space group \textit{I}4/\textit{mmm}. There is $\Gamma_1$, a two-dimensional irreducible representation with basis vector $(M_x,M_y,0)$, and $\Gamma_2$, a one-dimensional irreducible representation with basis vector $(0,0,M_z)$. Both representations do not have selection rules, in agreement with our observations, and hence can not be distinguished in this way.

We have determined that the structure must be described by $\Gamma_1$ with moments in the $ab$ plane by performing azimuthal scans. These scans measure the intensity variation when the sample is rotated by azimuthal angle $\Psi$ around the scattering vector $\bm{Q}$. In MRXD, the scattering intensity is  proportional to $| \bm{F}(\bm{Q}) \cdot \bm{k_f} | ^2$ where $\bm{F}(\bm{Q})$ is the magnetic structure factor and $\bm{k_f}$ is the scattered photon wavevector.\cite{Hill1996} Azimuthal rotations change the moment direction, modifying $\bm{F}(\bm{Q})$ relative to a fixed $\bm{k_f}$. The scattered intensity is therefore  expected to change with $\Psi$ and this can be compared with that expected according to a magnetic structure model. The azimuth $\Psi$ is defined relative to a reference Bragg peak, here chosen to be  $\bm{Q}~=~(-1,-1,0)$. The azimuthal angle is defined to be zero when the reference Bragg peak is in the scattering plane and forms the smallest angle with the incident photon wavevector $\bm{k_i}$. 

\begin{figure}[!htb]
\centering
\includegraphics[scale=0.55]{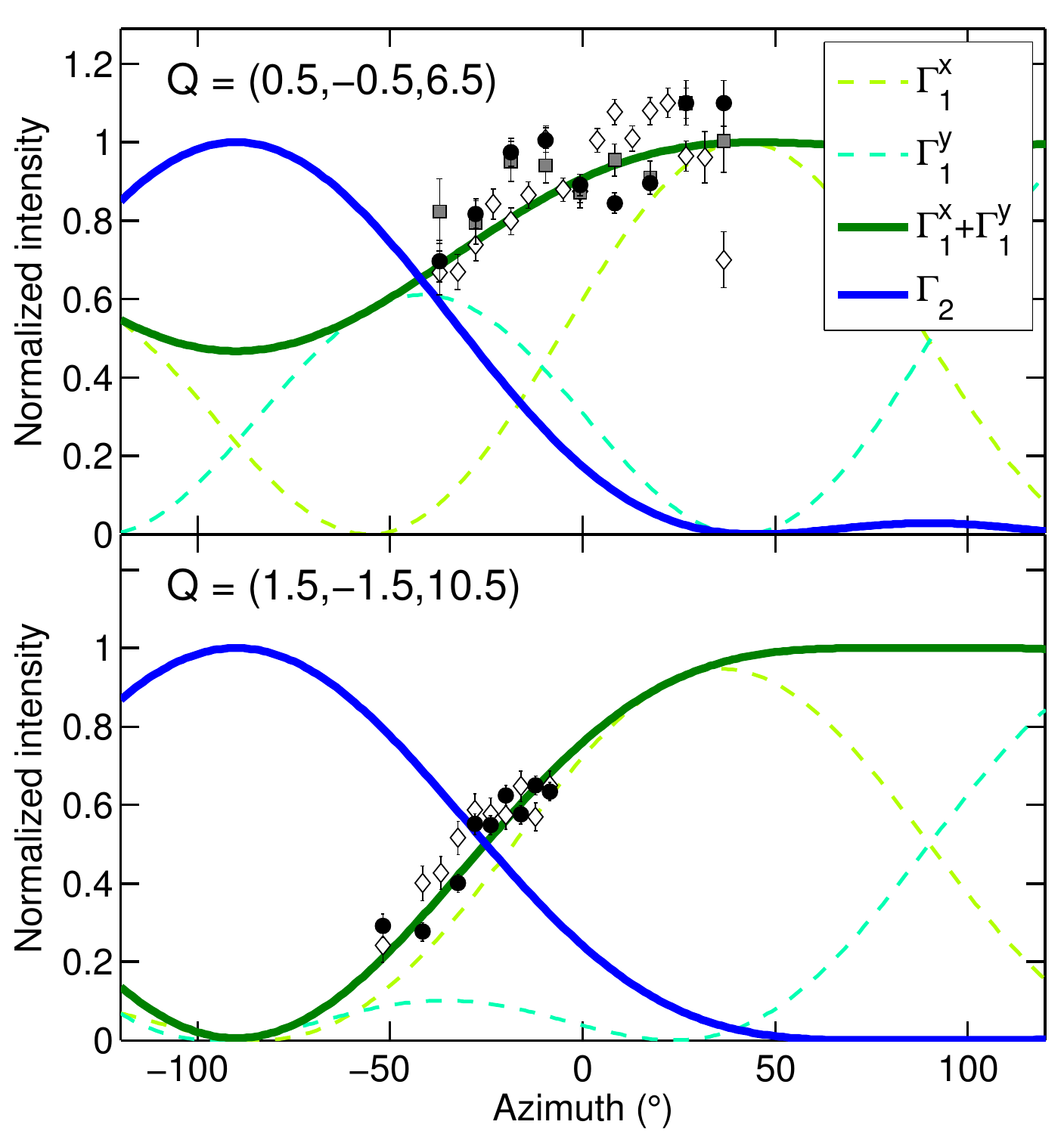}
\caption{Azimuthal scans on the magnetic Bragg peaks $\bm{Q}~=~(0.5, -0.5, 6.5)$ and $\bm{Q}~=~(1.5, -1.5, 10.5)$. The integrated intensities were corrected for absorption and scaled to compare to the models. $\Gamma_{1}^x$ and $\Gamma_{1}^y$ represent two domains with moments along $a$ and $b$ respectively. Assuming equal population of both domains, the model agrees well with the results.  $\Gamma_{2}$ represent the magnetic structure with moments along the $c$-axis. Multiple datasets are represented by different symbols (see text).}
\label{fig3}
\end{figure}

Multiple datasets of azimuthal scans were collected and are represented by different symbols in Fig.~\ref{fig3}. These datasets were collected in similar conditions (with and without optimizing the different rotation and translation motors) and all show the same general tendency. For both irreducible representations, the magnetic structure is collinear and the azimuthal scans correlate directly to the moment orientation. The theoretical azimuthal dependence curves for moments pointing along the $a$-axis ($\Gamma_1^x$), the $b$-axis ($\Gamma_1^y$) and the $c$-axis ($\Gamma_2$) are shown in Fig.~\ref{fig3} for the magnetic Bragg peaks $\bm{Q}~=~(0.5, -0.5, 6.5)$ and $\bm{Q}~=~(1.5, -1.5, 10.5)$. Experimental results are overlaid and show that the system can be described by the coexistence of $\Gamma_1^x$ and $\Gamma_1^y$ domains with equal population. Since the axes $a$ and $b$ are equivalent, one would indeed expect that both domains are present. In general, if a domain exists with a moment pointing in a direction $\bm{e}$ within the $ab$ plane, a domain with a moment pointing in a direction $\bm{e'}$ perpendicular to $\bm{e}$ in the $ab$ plane is expected with an equal population. It can be shown that the azimuthal dependence of $\Gamma_1^{\bm{e}}+\Gamma_1^{\bm{e'}}$ for any $\bm{e}$ in the $ab$ plane is exactly the same one as the one of $\Gamma_1^x+\Gamma_1^y$. Therefore, our results indicate the moments are in the $ab$ plane but do not allow us to determine their exact orientation. The magnetic structure for antiferromagnetically ordered moments pointing in the basal plane at an angle $\theta$ from the $a$-axis is schematized in Fig.~\refsub[d]{fig1}. 

From previous NQR experiments, it was claimed 
that at $T = 1.6$~K an incommensurate magnetic order coexists with the commensurate order, and that the volume fraction of commensurate:incommensurate order was 0.25:0.75.
In addition the maximal internal field due to the incommensurate order is determined to be slightly larger than the one from the commensurate order, suggesting a similar moment size for both orders. For these two reasons, the magnetic peak intensities originating from the incommensurate order can be expected to be similar to the ones of the commensurate order. However, no evidence for incommensurate magnetic peaks was found in our MRXD experiment from scans along the high symmetry directions in reciprocal space. Measurements were carried out at $T~=~1.8$~K for $\bm{Q}~=~(0.5,-0.5,L)$ from $L~=~6$ to 8, $\bm{Q}~=~(H,\overline{H},7)$ from $H~=~0$ to 1.2, $\bm{Q}~=~(H,\overline{H},6.5)$ from $H~=~0$ to 1.2 and $\bm{Q}~=~(H,0,6.5)$ from $H~=~0$ to 1.5. This rules out likely incommensurate propagation vectors similar to those of other incommensurate magnetic phases in Ce-based heavy-fermion compounds,\cite{Fobes2017,Christianson2005,Yokoyama2006,Ohira-Kawamura2007,Ohira-Kawamura2009,Raymond2014} but we can not exclude the presence of incommensurate modulations propagating elsewhere in reciprocal space. 

\begin{figure}[!tb]
\centering
\includegraphics[scale=0.6]{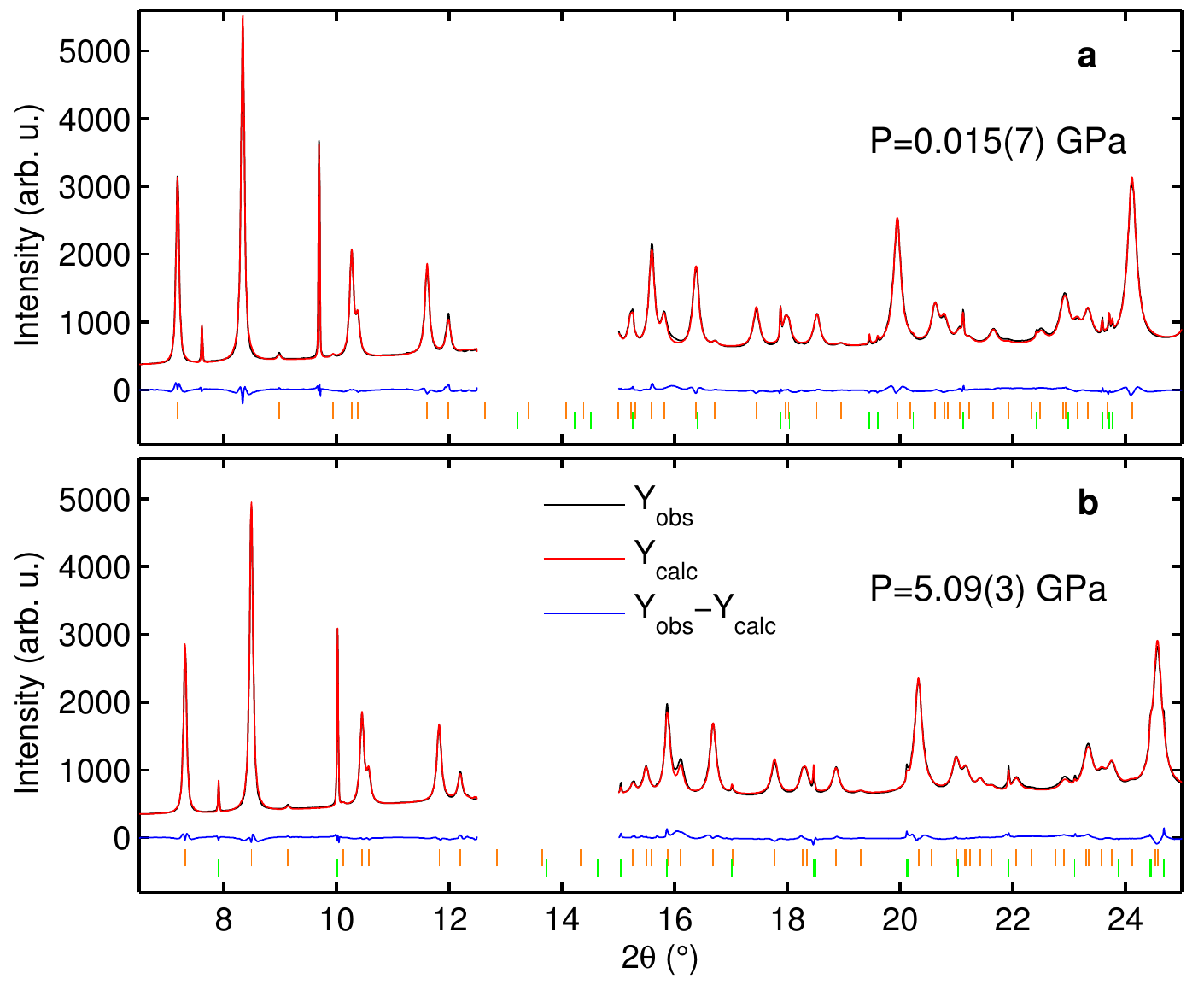}
\caption{X-ray diffraction pattern of \cpi\ at $T~=~293$~K for hydrostatic pressure of (a) $P~=~0.015(7)$~GPa and (b)~$P~=~5.09(3)$~GPa. The region $12.5^\circ<2\theta<15^\circ$ has been excluded from the refinement. Orange and green tick marks indicate the Bragg peak positions for \cpi\ and quartz, respectively. }
\label{fig4}
\end{figure}

\subsection{Powder X-ray Diffraction Under Pressure}

\begin{figure*}[ht]
\centering
\includegraphics[scale=1.1]{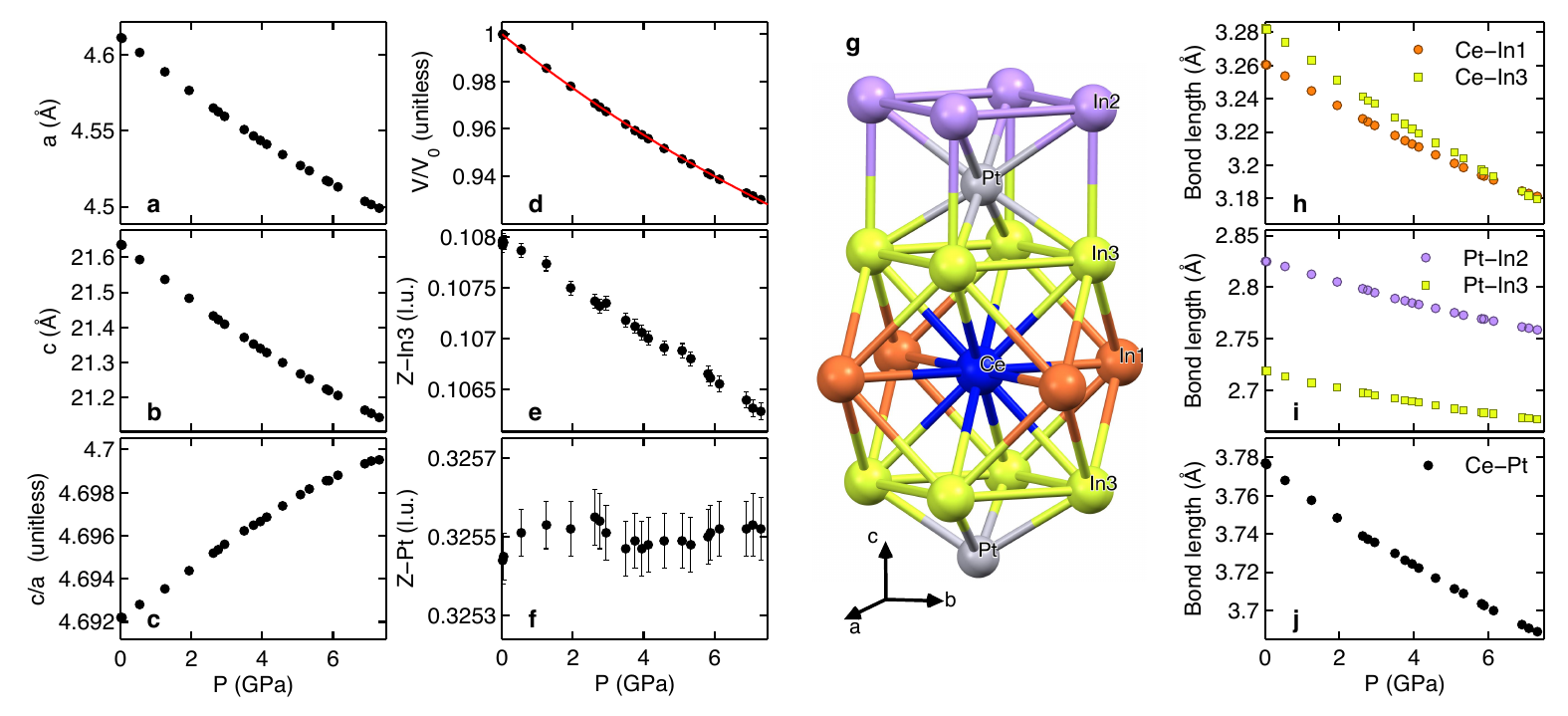}
\caption{Pressure dependence of the crystallographic structure at $T~=~293$~K. (a)-(c) Lattice constants $a$ and $c$ and ratio $c/a$ as function of pressure. (d) Unit cell volume as function of pressure, with the fit to the Birch-Murnaghan equation-of-state represented by the red solid line. (e)-(f)~Pressure dependence of the fractional coordinate $Z$ of Pt and In(3) atoms. (g) Crystallographic structure of \cpi\ around the Ce and Pt atoms, where In atoms are coloured by their site symmetry. (h)-(j) Pressure dependence of different bond lengths.}
\label{fig5}
\end{figure*}

Powder X-ray diffraction patterns of \cpi\ are shown in Fig.~\ref{fig4} at hydrostatic pressures $P~=~0.015(7)$~GPa and $P~=~5.09(3)$~GPa for a representative $2\theta$ angular range. The general crystallographic structure, previously reported by Klimczuk \textit{et al.},\cite{Klimczuk2014} was confirmed by Rietveld refinement using \textsc{FullProf}.\cite{Fullprof} Two strong diffraction peaks from \cpi\ appear in the angular range $12.5^\circ<2\theta<15^\circ$ and this region has been excluded from the refinement to improve the sensitivity of the fit to weak features over the full angular range. Importantly, the diffraction peak profiles due to the \cpi\ sample are significantly broader than the instrumental resolution and this can be attributed to strain. The presence of strain in polycrystalline sample of \cpi\ was inferred previously in NQR measurements.\cite{Sakai2011} Our measurements remained in a hydrostatic regime up to the maximal applied pressure, as confirmed by the pressure independent widths of peaks due to scattering from quartz. However, the peak widths of \cpi\ gradually broadened above $P\approx5$~GPa, which show a loss of the structural integrity in terms of either a larger strain or breaking of crystallites into smaller particles.

The refinement of the diffraction patterns was performed sequentially for increasing pressure and the results are presented in Fig.~\ref{fig5}. We observe no obvious changes of the crystallographic structure related to the characteristic pressures $P^*~=~2.4$~GPa and $P_c\approx3.4$~GPa. The lattice constants $a$ and $c$ change monotonically up to the maximal applied pressure $P~=~7.3$~GPa [Figs.~\refsub[a]{fig5}-\refsub[c]{fig5}]. The Birch-Murnaghan equation of state was used to relate the crystal volume $V$ to the applied pressure $P$:
\begin{equation}
P(V)=\frac{3}{2}B_0\left( v^7 - v^5\right)\left(1-\frac{3}{4}\left(4-B_0'(v^3-1\right)\right),
\end{equation}
where $B_0$, and $B_0'$ are respectively the initial bulk modulus and its derivative, and $v~=~(V_0/V)^{1/3}$.\cite{Birch1938} By fitting this equation to the data shown in Fig.~\refsub[d]{fig5}, we obtain $B_0~=~81.1\pm0.3$~GPa and $B_0'~=~5.8\pm0.1$. Using the simple Murnaghan equation~\cite{Murnaghan1937} results in the same fitted values for $B_0$ and $B_0'$ within errors. These values are similar to those reported for other members of the Ce$_n$M$_m$In$_{3n+2m}$ family.\cite{Kumar2004} In these compounds, it was observed that adding \textit{M}In$_2$ layers stiffens the structure and increases the bulk modulus: $B_0~=~67$~GPa for CeIn$_3$, average $B_0~=~70.4$~GPa for Ce$_2$\textit{M}In$_8$ (2 layers CeIn$_3$ + 1 layer \textit{M}In$_2$) and average $B_0~=~81.4$~GPa for Ce\textit{M}In$_5$ (1 layer CeIn$_3$ + 1 layer \textit{M}In$_2$). The addition of a second \textit{M}In$_2$ layer in \cpi\ relative to Ce\textit{M}In$_5$ could then be expected to stiffen the lattice further. However, the bulk moduli appear very similar for \cpi\ and the Ce-115s.

In \cpi, the Ce and Pt atoms sit at Wyckoff positions $2b$ and $4e$, respectively, and the In atoms are distributed on three different positions (In(1) at $2a$, In(2) at $4d$ and In(3) at $8g$). The only adjustable fractional coordinates in the structure of \cpi\ are the $Z$ positions of the Pt and In(3) atoms. The fractional coordinate $Z$ of In(3) changes monotonically with pressure [Fig.~\refsub[e]{fig5}] and the one of Pt is pressure independent [Fig.~\refsub[f]{fig5}]. This indicates a non-uniform compression along the $c$-axis, with the strongest contraction occurring between the In(3)-planes and the Ce-In(1) planes [see Fig.~\refsub[g]{fig5}]. The pressure dependence of various bond lengths is presented in Figs.~\refsub[h]{fig5}-\refsub[j]{fig5} and they all decrease monotonically with increasing pressure. Interestingly, the Ce-In(3) bond is more significantly affected by pressure than the Ce-In(1) bond [Fig.~\refsub[h]{fig5}]. 
Since the Ce-In coupling is expected to be the strongest with the out-of-plane In(3) atoms,\cite{Haule2010,Shim2007}
this change in distortion around the Ce atoms could modify significantly the ground-state Ce wavefunction.\cite{Willers2015} 

%%%%%%%%%%%%%%%%%%%%%%%%%%%%%%%%%%%%%%%%%%%%
%%%% Discussion 
%%%%%%%%%%%%%%%%%%%%%%%%%%%%%%%%%%%%%%%%%%%%
\section{Discussion}

As mentioned previously, the pressure-temperature phase diagram of CeRhIn$_5$ is very similar to the one of \cpi. Their magnetic structures at ambient pressure also share similarities: both have an antiferromagnetic order in the basal plane with moments lying in that plane.\cite{Fobes2017} However, the ordering in CeRhIn$_5$ is incommensurate along the $c$-axis in contrast with the commensurate ordering in \cpi. While CeCoIn$_5$ and CeIrIn$_5$ do not order magnetically at ambient pressure and zero magnetic field, it is possible to induce magnetic order with doping. In particular, substituting the Co or Ir sites with Rh leads to the coexistence of an incommensurate order with $\bm{k}~=~(\frac{1}{2},\frac{1}{2},\delta)$ and a commensurate order with $\bm{k}~=~(\frac{1}{2},\frac{1}{2},\frac{1}{2})$ for a range of doping values.\cite{Christianson2005,Yokoyama2006,Ohira-Kawamura2007,Ohira-Kawamura2009} It was shown for CeRh$_{0.7}$Ir$_{0.3}$In$_5$ specifically that the moments lie in the basal plane for both the commensurate and incommensurate orders. Doping the In site with Cd in CeCoIn$_5$ also stabilizes a commensurate order with $\bm{k}~=~(\frac{1}{2},\frac{1}{2},\frac{1}{2})$.\cite{Nicklas2007} On the other hand, substituting Ce by Nd in CeCoIn$_5$ leads to a propagation vector $\bm{k}~=~(\frac{1}{2}-\delta,\frac{1}{2}-\delta,\frac{1}{2})$ with $\delta~=~0.05$,\cite{Raymond2014} suggesting a spin-density wave in the basal plane with fundamentally different properties from the localized moment magnetism in CeRhIn$_5$ and \cpi.

In these systems, superconductivity emerges in the vicinity of an AFM QCP, suggesting a magnetically-driven pairing mechanism of superconductivity. The knowledge of the magnetic structure is therefore a crucial element for identifying the magnetic fluctuations responsible for this electron-electron coupling. The AFM order $(\frac{1}{2},\frac{1}{2})$ in the basal plane prevails in these systems and \cpi\ appears as a new example where magnetic fluctuations associated with this AFM order are the pairing glue of the pressure-induced superconductivity. It is important to note that the magnetic structure of \cpi\ might change under pressure but it is unlikely to change the order in the basal plane. For example, the propagation vector in CeRhIn$_5$ changes under pressure but the order in the basal plane is conserved.\cite{Majumdar2002,Llobet2004,Raymond2008,Aso2009}

Based on NQR experiments, it was suggested that in single crystals of \cpi\ there is a coexistence of commensurate and incommensurate orders at ambient pressure.\cite{Sakai2014,Sakai2011} Specifically, sharp peaks in the spectrum can be attributed to a basal plane AFM order with moments pointing along the $a$-axis or the $b$-axis. This was interpreted as a commensurate order. On the other hand, broad features are also observed in the spectrum and were attributed to a distribution of internal fields at the In(2) and In(3) sites. This was interpreted as an incommensurate order similar to the one of CeRhIn$_5$.\cite{Curro2006} 

Our results presented in section~\ref{secMRXD} confirm the presence of a commensurate order but {do not reveal the presence of an} incommensurate order along {the} high symmetry direction{s} in reciprocal space {indicated in section~\ref{secMRXD}}. {The scenario involving the coexistence of both commensurate and incommensurate orders remains a possibility: we cannot rule out incommensurate modulations propagating elsewhere in reciprocal space, and the volume fraction and/or moment size could be too small to be detected under our current experimental conditions.} 

{On the other hand, we} propose an alternative interpretation of the broad features observed in the NQR experiments that do not require the presence of an additional incommensurate order. 
With no restriction on the precise moment direction in the basal plane provided by our MXRD experiments, the distribution of internal fields observed by NQR could be generated if either the moment directions in the $ab$ plane fluctuate, or there exist multiple domains with different moment orientations (different values of $\theta$ in Fig.~\ref{fig1}).
This commensurate-only scenario for the magnetic order in \cpi\ requires a coexistence of domain-types; those with arbitrary moment orientations in the $ab$ plane as outlined above, and those where the moments are rigidly aligned with the $a$- and $b$-axes. Here crystal strain could play an important role in stabilizing one type of domain over the other.

In NQR experiments, different results for the reported spectra are obtained from polycrystalline and single crystal samples of \cpi.\cite{Sakai2011,Aproberts-Warren2010} These discrepancies are readily attributable to crystal/surface strain effects that vary in propensity with the sample crystallite size. Indeed, this is supported by the broad structural peaks in our high-resolution powder X-ray diffraction experiment on \cpi. In the NQR studies only sharp features are observed for powder samples, in contrast with the presence of broad features for single crystals. Furthermore, applied pressure on single crystals suppresses the contribution of the broad features.\cite{Sakai2011} Taken together, these two effects indicate that strain, either from surface strain from the grains in polycrystalline samples or stimulated by pressure, promotes the ordering with moments aligned along the $a$-axis or the $b$-axis. At the same time, in the absence of strain, the moments may align along an arbitrary direction in the $ab$ plane. In this scenario, enhanced strain thus leads to an effective in-plane anisotropy that favors the alignment of the moments along the $a$-axis or $b$-axis.

It is interesting then to note that the superconductivity is stabilized in a wider pressure range in powder samples and that it only appears in single crystals when the NQR signature interpreted in terms of incommensurate order is completely suppressed.\cite{Sakai2011,Sidorov2013} This suggests that domains with moments not aligned along the $a$-axis or the $b$-axis are detrimental to the formation of superconductivity in \cpi.

Finally, we note that even if the magnetic structure presented in section~\ref{secMRXD} is the simplest solution to explain the results, it is not the only possible one. Since the lattice of \cpi\ is body-centered, the propagation vector $\bm{k}_{1/2}$ is not equivalent to $-\bm{k}_{1/2}$. This can lead either to two different $\bm{k}$-domains, which was assumed in section~\ref{secMRXD}, or a multi-$\bm{k}$ structure, as observed for example in the heavy fermion CeRh$_2$Si$_2$, which also has a body-centered tetragonal lattice.\cite{Kawarazaki2000} A complete description of the multi-$\bm{k}$ structure in \cpi\ is given in the appendix. In such a multi-$\bm{k}$ structure, the moments between the nearest neighbouring Ce layers can be non-collinear while all the moments are collinear in a single-$\bm{k}$ structure. This non-collinearity suggests an effective decoupling of the nearest neighbour layers while keeping a coupling to the next-nearest neighbour planes, consequently forming two decoupled yet inter-penetrating sublattices. This scenario is plausible for the body-centered tetragonal lattice because of the presence of competing interactions. It was even suggested theoretically that the frustration in body-centered tetragonal lattices can destabilize long-range magnetic order and lead to spin liquid states in heavy fermion compounds.\cite{Farias2016}
The aforementioned discussion about the moment directions in the single-$\bm{k}$ model, and its application for consistently explaining previously reported NQR spectra, can also be done using the multi-$\bm{k}$ structure. Our results do not allow us to establish unambiguously if the single-$\bm{k}$ structure or the multi-$\bm{k}$ structure is the correct one. In fact, these two scenarios cannot be distinguished in a simple scattering experiment; doing so would require the application of either uniaxial strain or magnetic fields to control the magnetic domain formation in a single crystal sample.

%%%%%%%%%%%%%%%%%%%%%%%%%%%%%%%%%%%%%%%%%%%%%
%%%%%%%%%%%%%%%%%%%%%%%%%%%%%%%%%%%%%%%%%%%%%
%%%%%%%               Conclusions
%%%%%%%%%%%%%%%%%%%%%%%%%%%%%%%%%%%%%%%%%%%%%
%%%%%%%%%%%%%%%%%%%%%%%%%%%%%%%%%%%%%%%%%%%%%
\section{Summary}

We have shown that the crystallographic structure of \cpi\ changes monotonically with pressure up to $P~=~7.3$~GPa at room temperature. We also investigated the magnetic order of \cpi\ at ambient pressure below $T_N~=~5.34(2)$~K by magnetic resonant X-ray diffraction. This order is characterized by a commensurate propagation vector $\bm{k}_{1/2}~=~\left( \frac{1}{2} , \frac{1}{2}, \frac{1}{2}\right)$. The magnetic origin of these diffraction peaks was confirmed by their resonance at the Ce-$L_{II}$ absorption edge and by polarization analysis. Azimuthal scans confirm that the moments lie in the basal plane. The magnetic structure can be described by a single-$\bm{k}$ structure or by a multi-$\bm{k}$ structure. Both structures cannot be distinguished in a simple scattering experiment as reported here and the single-$\bm{k}$ structure is discussed for simplicity. The presence of incommensurate order in \cpi\ was previously reported based on NQR experiments. {Our measurements could not reveal the presence of such an order but are insufficient to exclude it completely.}
Using our results we propose a new scenario for the ambient pressure ground state of \cpi\ that is described only by commensurate magnetic order; namely a coexistence of domains wherein the moments are either rigidly aligned along the $a$- and $b$-axes, or arbitrarily aligned within the $ab$ plane. Crystal strain is argued to be an effective tuning parameter for controlling the relative volume fractions of the two types of domain, thus providing a means for a consistent description of both the scattering data reported here, and previously reported NQR spectra obtained on both polycrystalline and single crystal samples.

\section*{Acknowledgements}

The authors are thankful to M. Kenzelmann, D. G. Mazzone and F. Ronning for fruitful discussions. This research received support from the Natural Sciences and Engineering Research Council of Canada (Canada). Work at Los Alamos National Laboratory was performed under the auspices of the US Department of Energy, Office of Basic Energy Sciences, Division of Materials Sciences and Engineering. XMaS is a UK mid-range facility supported by EPSRC.

\textit{Note added.} - During the preparation of this manuscript, we became aware of another report where the magnetic structure of \cpi\ was investigated using neutron diffraction.\cite{Raba2017} In agreement with our results, they report a commensurate propagation vector $\bm{k}_{1/2}$ and moments lying in the basal plane. The reported structure corresponds to a multi-$\bm{k}$ structure with non-collinear moments.

%%%%%%%%%%%%%%
\appendix*
\section{{Single-$k$ and multi-$k$ structures}}

\subsection{Single-$k$ structure}

\begin{figure*}[!htb]
\centering
\includegraphics[scale=0.9]{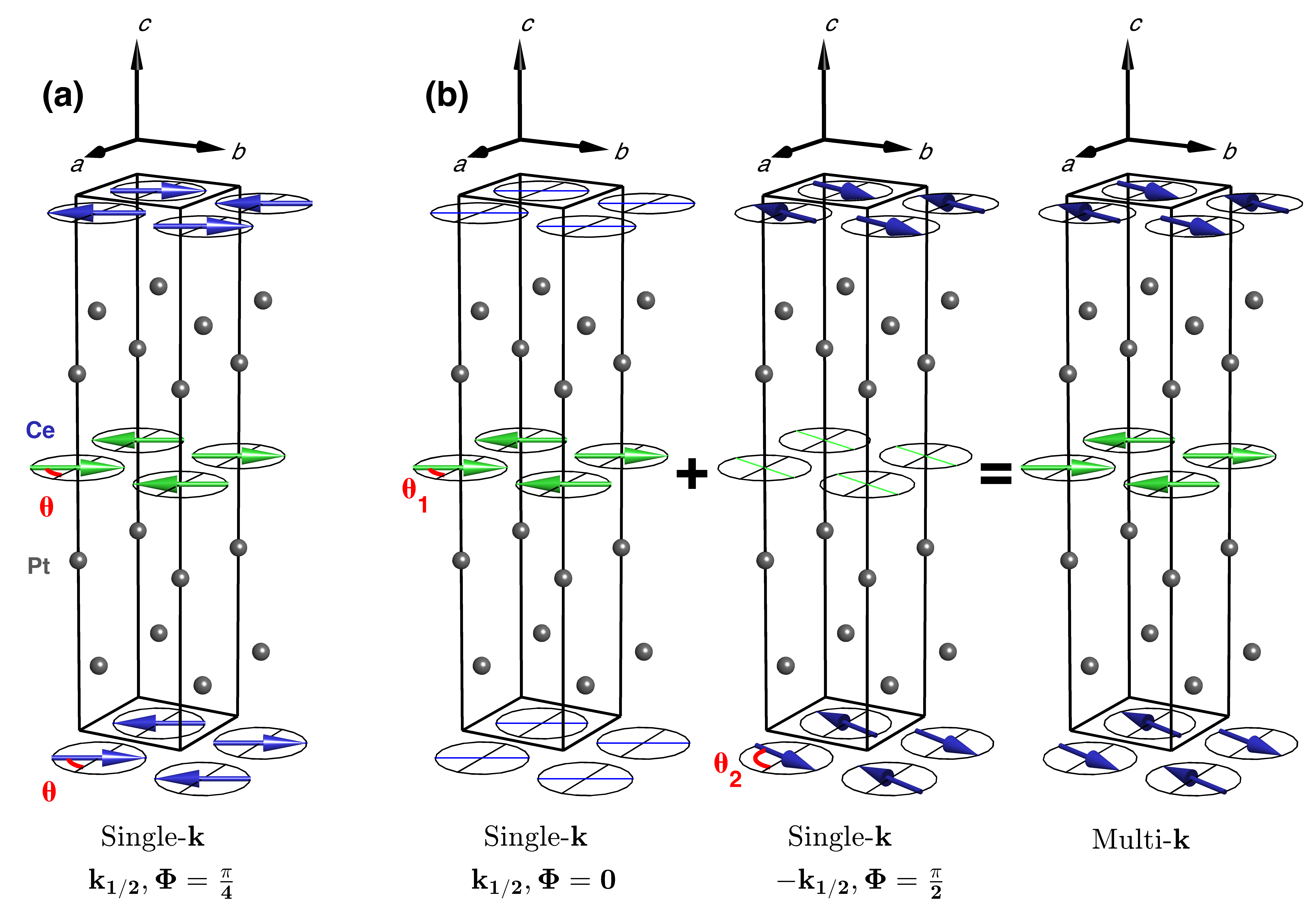}
\caption{(a) Single-$\bm{k}$ structure with the propagation vector $\bm{k}_{1/2}$ and the phase factor $\Phi~=~\frac{\pi}{4}$. (b) Single-$\bm{k}$ structures with $\bm{k}~=~\bm{k}_{1/2}$, $\Phi~=~0$ and $\bm{k}~=~-\bm{k}_{1/2}$, $\Phi~=~\frac{\pi}{2}$, combined to form a multi-$\bm{k}$ structure. The angles $\theta_1$ and $\theta_2$ indicate the moment direction in the basal plane relative to the $a$-axis for each single-$\bm{k}$ structure.}
\label{fig6}
\end{figure*}

The simplest magnetic structure model for \cpi\ is described by the single propagation vector $\bm{k}_{1/2}$. This is represented in Fig.~\refsub[a]{fig6}. In the unit cell, there are two Ce ions which are related by the body-centering symmetry. They are distinguished by the blue and green colors in Fig.~\ref{fig6}. For a general single-\textbf{k} structure, the moments $m_b$ and $m_g$ at the blue and green sites, respectively, are expressed as:
\begin{equation}
 {\bm{m}}_b = 
  \begin{pmatrix} 
 M \cos \theta \\
 M \sin \theta \\
 0
    \end{pmatrix}
  \cos \Phi 
\end{equation}
\begin{equation}
 {\bm{m}}_g = 
  \begin{pmatrix} 
 M \cos \theta \\
 M \sin \theta \\
 0
    \end{pmatrix}
  \cos (\Phi\pm\frac{\pi}{2}).
\end{equation}
Here the parameter $\theta$ is the angle of the moment in the $ab$ plane, which can take any value. To reproduce the data, the presence of two equally populated domains with $\theta$ and $\theta+90^\circ$ is assumed. The parameter $\Phi$ is a global phase that cannot be measured with scattering techniques. For physical reasons, we chose $\Phi=\frac{\pi}{4}$ to generate equal moments for $ {\bm{m}}_b$ and $ {\bm{m}}_g$. The single-$\bm{k}$ structure is therefore defined by:
\begin{equation}
 {\bm{m}}_b =  \frac{1}{\sqrt{2}}
  \begin{pmatrix} 
 M \cos \theta \\
 M \sin \theta \\
 0
    \end{pmatrix}
\end{equation}
\begin{equation}
 {\bm{m}}_g = - \frac{1}{\sqrt{2}}
  \begin{pmatrix} 
 M \cos \theta \\
 M \sin \theta \\
 0
    \end{pmatrix}.
\end{equation}

%%%%%%%%%%%%%%%%%%%%%%%%%%%%%%%%%%%%%%%%%%%
%%%%%%%%%%%%%%%%%%%%%%%%%%%%%%%%%%%%%%%%%%%
%%%%%%%%%%%%%%%%%%%%%%%%%%%%%%%%%%%%%%%%%%%
%%%%%%%%%%%%%%%%%%%%%%%%%%%%%%%%%%%%%%%%%%%
\subsection{Multi-$k$ structure}
Due to the body-centering symmetry, $+\bm{k}_{1/2}$ and $-\bm{k}_{1/2}$ are not equivalent and therefore, a magnetic structure can form that is composed of two propagation vectors. In a general way, the moments are defined at the blue and green sites, respectively, by:
\begin{equation}
 {\bm{m}}_b^{+\bm{k}} +  {\bm{m}}_b^{-\bm{k}} = 
  \begin{pmatrix} 
 M \cos \theta_1 \\
 M \sin \theta_1 \\
 0
    \end{pmatrix}
  \cos \Phi_1
  + 
   \begin{pmatrix} 
 M \cos \theta_2 \\
 M \sin \theta_2 \\
 0
    \end{pmatrix}
  \cos \Phi_2
\end{equation}
\begin{equation}
\begin{split}
 {\bm{m}}_g^{+\bm{k}}  +  {\bm{m}}_g^{-\bm{k}} & = 
  \begin{pmatrix} 
 M \cos \theta_1 \\
 M \sin \theta_1 \\
 0
    \end{pmatrix}
  \cos (\Phi_1 + \frac{\pi}{2})
  \\ & + 
    \begin{pmatrix} 
   M \cos \theta_2 \\
 M \sin \theta_2 \\
 0
    \end{pmatrix}
  \cos (\Phi_2 - \frac{\pi}{2})
  \end{split}
\end{equation}
where $\theta_1$ and $\Phi_1$ are related to the propagation $+\bm{k}_{1/2}$, and $\theta_2$ and $\Phi_2$ are related to $-\bm{k}_{1/2}$. It is again assumed that there are two equally populated domains with $\{\theta_1,\theta_2 \}$ and $\{\theta_1  +90^\circ,\theta_2 +90^\circ \}$. Experimentally, this gives exactly the same scattering as the single-$\bm{k}$ structure. We must choose $\Phi_1$ and $\Phi_2$ to have equal moments on the blue and green sites for any $\theta_1$ and $\theta_2$. An elegant choice is $\Phi_1=\frac{n \pi}{2}$ and $\Phi_2=\frac{(n+1) \pi}{2}$ where $n$ is an integer. It evidences the decoupling of the nearest-neighbour layers. For $n=0$, we obtain:
\begin{equation}
 {\bm{m}}_b^{+\bm{k}} +  {\bm{m}}_b^{-\bm{k}} = 
  \begin{pmatrix} 
 M \cos \theta_1 \\
 M \sin \theta_1 \\
 0
    \end{pmatrix}
\end{equation}
and
\begin{equation}
 {\bm{m}}_g^{+\bm{k}}  +  {\bm{m}}_g^{-\bm{k}} = 
  \begin{pmatrix} 
   M \cos \theta_2 \\
 M \sin \theta_2 \\
 0
    \end{pmatrix}.
\end{equation}
The structure is therefore defined by three parameters: the moment size $M$, the angle $\theta_1$ of the first propagation vector and the angle $\theta_2$ of the second propagation vector. While $M$ is expected to be constant, $\theta_1$ and $\theta_2$ can take any value. Note that the single-$\bm{k}$ structure is obtained if $\theta=\theta_1=\theta_2$.

\bibliographystyle{apsrev4-1}
%\input{CePt2In7_170717_biblio}
%\bibliography{biblioJuly2017v2}


\end{document}